# Are Defect Profile Similarity Criteria Different Than Velocity Profile Similarity Criteria for the Turbulent Boundary Layer?

David Weyburne[1]
Air Force Research Laboratory
2241 Avionics Circle
WPAFB OH 45433

The use of the defect profile instead of the experimentally observed velocity profile for the search for similarity parameters has become firmly imbedded in the turbulent boundary layer literature. However, a search of the literature reveals that there are no theoretical reasons for this defect profile preference over the more traditional velocity profile. In the report herein, we use the flow governing equation approach to develop similarity criteria for the two profiles. Results show that the derived similarity criteria are identical. Together with previous work that found that defect profile similarity must be accompanied by velocity profile similarity, then ones expectations must be that either profile can be used to search for similarity in experimental datasets. The choice should therefore be dictated by which one works best for experimental investigations, which in this case is the velocity profile.

## 1. Introduction

One of the most fundament concepts in fluid mechanics research is the search for ways to scale experiment observables so that the scaled observable from different measurement stations along the flow appear to be similar. Similarity of the velocity profile formed by fluid flow along a wall is one of those much studied fundamental cases. For 2-D wall-bounded flows, velocity profile similarity is defined as the case where two velocity profiles taken at different stations along the flow differ only by simple scaling parameters in $y$ and $u(x,y)$, where $y$ is the normal direction to the wall, $x$ is the flow direction, and $u(x,y)$ is the velocity parallel to the wall in the flow direction. Similarity solutions to the flow governing equations are well known for laminar flow. Turbulent flow similarity is more problematic. Since the equations for turbulent flows do not admit to exact similarity solutions, the community has sought to establish their possible existence by looking for scaling parameters that collapse experimental velocity profile datasets to a single curve. The early search for similarity scaling behavior for the turbulent boundary layer was energized by the experimental and theoretical work of Clauser [1]. In his similarity investigation, Clauser found that datasets plotted as defect profiles, defined as $u_e - u(x,y)$ where $u_e$ is the velocity at the boundary layer edge, resulted in similarity behavior whereas the same data plotted as velocity profiles did not. Subsequent theoretical approaches to turbulent boundary layer similarity by Rotta [2] and Towsend [3] followed Clauser and pursued a theoretical approach to similarity of the defect profile. This preference for the use of the defect profile is also deeply impeded in the Logarithmic Law of the Wall theory for the turbulent boundary layer. It has its roots in von Kármán's [4] development of the analytical expression of the defect profile for wall-bounded turbulent boundary layers. Taken together, the use of the defect profile has become the accepted method for the describing/discussing similarity of the turbulent boundary layer.

[1]Email: David.Weyburne@us.af.mil

However, there is a problem here. To accept the Clauser result showing defect profile similarity but not velocity profile similarity one is making the assumption that the underlying physics of the defect profile is different and distinct from that of the velocity profile. The problem with this assumption is that the defect profile is a simple calculated profile that is based solely on the experimental velocity profile data taken at that measurement location (note that $u_e$ is calculated from the velocity profile data). The distinction assumption implies that somehow one is changing the physics of the flow merely by shifting the data and presenting it in a different way. Furthermore, a search of the literature reveals that there are no theoretical justifications for this distinction, *i.e*. it has never been shown that defect profile similarity is different and distinct from velocity profile similarity. This distinction assumption would imply, for example, that the defect profile similarity scaling parameters are different than the similarity scaling parameters for the velocity profile. If this is in fact the case then one should expect that the similarity criteria that are derived for the two profiles should be different and distinct. Herein we use a flow governing equation approach to similarity to show that in fact the derived similarity criteria for the two cases are identical. Hence, it must be the case that similarity scaling parameters should work equally well for the defect profile as well as the velocity profile.

If in fact there is no distinction between the defect profile and the velocity profile in terms of similarity criteria then how can one explain the results observed by Clauser? In that paper, Clauser [1] took datasets from the literature and plotted them as either velocity profiles (his Fig. 2) or defect profiles (his Fig. 3). The defect profiles showed similar behavior but the velocity profiles did not when both sets were scaled with the skin friction velocity and the boundary layer thickness. Recently Weyburne [5] observed this same type of behavior in other datasets taken from the literature that were used to claim similar behavior based on plots of the defect profile. After further investigation, Weyburne found that those datasets were cases of what he termed "false" similarity. The difference between what Weyburne called true similarity and false similarity is that in the false similarity case, the defect profiles appear to show similarity but one of the key similarity criteria is not satisfied. The tip-off to the existence of this false similarity scenario is exactly what Clauser and Weyburne observed; the dataset will show defect profile similarity but not velocity profile similarity. For the sake of completeness, this false similarity concept is described in detail below. But first, the similarity criteria for the defect profile is calculated and compared to velocity profile similarity criteria.

## 2. Flow Governing Equation Development

The use of the defect profile in similarity studies has associated with it an assumption that the defect profile is distinct from the velocity profile. This implies that it is possible to have defect profile scaling parameters that are different then the scaling parameters for the velocity profile. If this is in fact the case then one should expect that the similarity criteria that are derived for the two profiles should be different and distinct. One way to develop similarity criteria is to use the flow governing equation approach. The flow governing equation approach to similarity for 2-D flow assumes that the stream-wise velocity $u(x,y)$ and the wall-normal velocity $v(x,y)$ can be expressed as a product of *x*-functionals and scaled *y*-functionals. These *x* and *y* functional products are then introduced into the flow governing equations in order to reduce the flow governing equations to a dimensionless set of equations. In this approach, defect profile similarity and velocity profile similarity amounts to making different stream function assumptions. So the question becomes; are the defect profile similarity criteria different then the velocity profile similarity criteria for the two flow governing equation approaches? If the scaling criteria are the same then it is not possible to have a length and velocity scaling parameter work for one case but not the other. Hence, our approach to investigating similarity differences is to compare the scaling criteria developed using the flow governing equation approach for the defect profile to those developed for the velocity profile.

Both approaches start by modeling the fluid flow past a plate theoretically by a combination of the Reynolds-averaged Navier-Stokes equation and the continuity equation. This starts by placing the *x*-axis in the plane of the plate in the flow direction, the *y*-axis is placed at right angles to this plate, and the *z*-axis is placed along the plane of the plate perpendicular to the flow direction. For a 2-D, incompressible turbulent boundary layer, the Prandtl boundary layer *x*-component of the momentum balance is given by

$$u\frac{\partial u}{\partial x} + v\frac{\partial u}{\partial y} + \frac{\partial}{\partial x}\left\{\overline{\tilde{u}^2} - \overline{\tilde{v}^2}\right\} + \frac{\partial}{\partial y}\left\{\overline{\tilde{u}\tilde{v}}\right\} \cong -\frac{1}{\rho}\frac{\partial p}{\partial x} + \nu\frac{\partial^2 u}{\partial y^2} \quad , \tag{2.1}$$

where $\rho$ is the density, $p$ is the pressure, and $\nu$ is the kinematic viscosity (see Appendix A for full details). The fluctuating components are $\tilde{u}$ and $\tilde{v}$. The process of reducing this equation to a dimensionless equation is given in exact detail in the Appendix A. In what follows we give a brief outline of the Appendix A formulation.

## 2.1 Traditional Stream Function and Variable Transform

In order to reduce the *x*-momentum equation to a dimensionless equation we introduce the dimensionless independent variables

$$\xi = C_\xi x \, , \quad \eta = \frac{y}{\delta_s(x)} \quad , \tag{2.2}$$

where $C_\xi$ is a constant with units of 1/distance and the function $\delta_s(x)$ is an as yet unknown boundary layer thickness scaling parameter that is only a function of *x*. Next we introduce a stream function so that the mass conservation equation will automatically be satisfied. Underlying this stream function approach is a critical assumption to the whole theoretical development and that is that the velocities $u(x,y)$ and $v(x,y)$ can be decomposed into a product of *x*-functionals and scaled *y*-functionals. Thus we assume that a stream function $\psi(x,y)$ exists such that

$$\psi(x,y) = \delta_s(x)u_s(x)f(\xi,\eta) \quad , \tag{2.3}$$

where $u_s(\mathrm{x})$ is the velocity scaling parameter, $f(\xi,\eta)$ is a dimensionless function, and that the stream function satisfies the conditions

$$u(x,y) = \frac{\partial\psi(x,y)}{\partial y}, \qquad v(x,y) = -\frac{\partial\psi(x,y)}{\partial x} \quad . \tag{2.4}$$

The reduction of the x-momentum equation (Eq. 2.1) to a dimensionless equation is very tedious so it is moved to Appendix A. The result of transforming the *x*-component of the momentum balance is given as Eq. A.15 and repeated here as

$$u_s \frac{du_s}{dx} f'^2 - \frac{u_s}{\delta_s} \frac{d\{\delta_s u_s\}}{dx} ff'' + u_s^2 f' \frac{\partial f'}{\partial \xi} - \delta_s u_s f'' \frac{\partial f}{\partial \xi} + \qquad (2.5)$$

$$+\{\text{Reynolds Stress Terms}\} = u_e \frac{du_e}{dx} + \nu \frac{u_s}{\delta_s^2} f''' \quad ,$$

where the prime indicates differentiation with respect to $\eta$. The Reynolds stress terms are left un-transformed since they can have no involvement with the behavior of $u(x,y)$ or $v(x,y)$ without invoking some type of closure assumption.

Recall that similarity of the velocity profile for 2-D wall-bounded flows is defined as the case where two velocity profiles taken from different stations along the flow differ only by simple scaling parameters in $y$ and $u(x,y)$. Hence these scaling parameters must only be a function of *x*. In order to see this behavior in Eq. 2.5, we start by assuming that the functional $f(\xi,\eta)$ only depends on the scaled *y*-parameter $\eta$. This means that the terms involving differentiation with respect to $\xi$ must be zero. Furthermore, similarity requires that each *x*-variable grouping must change proportionately as we move from one measurement station to the next. Alternatively, we can divide Eq. 2.5 through by one of the *x*-variable groupings which then changes the similarity condition to be that each *x*-variable grouping ratio must be constant. Using the latter approach, we divide Eq. 2.5 by $u_s\, du_s/dx$. Thus the pressure gradient *x*-grouping ratio becomes

$$\frac{u_e \dfrac{du_e}{dx}}{u_s \dfrac{du_s}{dx}} = \frac{1}{C_0^2} \quad , \qquad (2.6)$$

where $C_0$ must be a constant for similarity (the reason for writing the constant term in this way will become obvious shortly). Let

$$\Lambda_s = -\frac{\delta_s}{u_s} \frac{du_s/dx}{d\delta_s/dx} \quad , \qquad (2.7)$$

and use this in the next *x*-grouping term given by

$$\frac{\dfrac{u_s}{\delta_s} \dfrac{d\{\delta_s u_s\}}{dx}}{u_s \dfrac{du_s}{dx}} = \frac{u_s \dfrac{du_s}{dx} + \dfrac{u_s^2}{\delta_s} \dfrac{d\delta_s}{dx}}{u_s \dfrac{du_s}{dx}} = 1 + \frac{\dfrac{u_s^2}{\delta_s} \dfrac{d\delta_s}{dx}}{u_s \dfrac{du_s}{dx}} = 1 + \frac{u_s \dfrac{d\delta_s}{dx}}{\delta_s \dfrac{du_s}{dx}} = 1 - \frac{1}{\Lambda_s} \quad . \qquad (2.8)$$

Hence $\Lambda_s$ must be a constant for similarity. The viscous *x*-grouping ratio is

$$\frac{\nu \dfrac{u_s}{\delta_s^2}}{u_s \dfrac{du_s}{dx}} = \frac{\nu}{\delta_s^2 \dfrac{du_s}{dx}} = \frac{1}{\beta} \quad , \qquad (2.9)$$

where $\beta$ must be a constant for similarity. Substituting into Eq. 2.5 we get the transformed *x*-component of the momentum balance as

$$\frac{f'''}{\beta}+\left(1-\frac{1}{\Lambda_s}\right)ff''+\frac{1}{C_0^2}-f'^2+\{\text{Reynolds Stress Terms}\} \quad = \quad 0 \quad , \tag{2.10}$$

$$f'''+\beta\left(1-\frac{1}{\Lambda_s}\right)ff''+\beta\left(\frac{1}{C_0^2}-f'^2\right)+\{\text{Reynolds Stress Terms}\} \quad = \quad 0 \quad ,$$

where we have assumed $f$ is now only a function of $\eta$.

The equation begins to take on the appearance of the familiar Falkner-Skan equation [11] as it should. To emphasize this we introduce $\alpha$ so that

$$\alpha \quad = \quad \beta\left(1-\frac{1}{\Lambda_s}\right) \quad = \quad \frac{\delta_s^2 \dfrac{du_s}{dx}}{\nu}\left(1+\frac{u_s \dfrac{d\delta_s}{dx}}{\delta_s \dfrac{du_s}{dx}}\right) \tag{2.11}$$

$$\alpha \quad = \quad \frac{\delta_s^2 \dfrac{du_s}{dx}}{\nu}+\frac{\delta_s u_s \dfrac{d\delta_s}{dx}}{\nu} \quad .$$

Substituting these parameters into Eq. 2.6, the dimensionless momentum equation becomes

$$f'''+\alpha ff''+\beta\left(\frac{1}{C_0^2}-f'^2\right)+\{\text{Reynolds Stress Terms}\} \quad = \quad 0 \quad . \tag{2.12}$$

## 2.1.a Velocity Profile Similarity Criteria

The *x*-grouping ratios given by Eqs. 2.6, 2.8, and 2.9 constitute the similarity criteria. Similarity requires that the three parameters $C_0$, $\Lambda_s$, and $\beta$ must be constant otherwise the solutions obtained at each station would be different. Of these three, only the $C_0$ derived from the pressure gradient *x*-grouping ratio involves both our unknown velocity scaling parameter $u_s(x)$ and the velocity at the boundary layer edge $u_e(x)$. We can write Eq. 2.7 as

$$u_s \frac{du_s}{dx} \quad = \quad C_0^2 u_e \frac{du_e}{dx} \quad , \tag{2.13}$$

where $C_0^2$ is a constant. This differential equation has a solution given by

$$u_s(x)=\sqrt{a+u_e^2(x)C_0^2} \quad , \tag{2.14}$$

where $a$ is a constant. There is no restrictions on the value of $a$ so we take $a$=0 which then means that one of the non-trivial solutions is given by

$$u_s(x)/u_e(x) \quad = \quad C_0 \quad . \tag{2.15}$$

This is one of the similarity conditions discovered by Rotta [2], Townsend [3], Castillo and George [6], Maciel, Rossignol, and Lemay [7], and Jones, Nickels, and Marusic [8] using defect profile flow governing equation approaches.

If the fluctuating components $\tilde{u}$ and $\tilde{v}$ are identically zero, then the Reynolds stress terms in our dimensionless momentum equation, Eq. 2.12, disappear. Couple that with taking $a$=0 and $C_0 = 1$ in Eq. 2.14, then Eq. 2.12 reduces to

$$f''' + \alpha f f'' + \beta\left(1 - f'^2\right) \;=\; 0 \quad , \tag{2.16}$$

which is the Falkner-Skan [11] similarity equation for laminar flow on a flat plate with a pressure gradient. In this case Eq. 2.16, and $\alpha$ and $\beta$ terms (Eqs. 2.9 and 2.11), are identical to Schlichting's [9] version of the Falkner-Skan equation. Taking $a$=0 and $C_0 = 1$ in Eq. 2.14 is consistent with the original laminar flow Falkner-Skan equation which assumes that $u_s(x)$=$u_e(x)$ right from the start.

## 2.2 Defect Stream Function and Variable Transform

Now as a reminder, the intent here in Section 2 is to show that the similarity conditions calculated using the defect profile are the same (or not) as the ones calculated for the velocity profile. In the last subsection we derived the similarity constraints for the velocity profile based equation. Now we turn our attention to the defect profile. Rotta [2], Townsend [3], Castillo and George [6], Maciel, Rossignol, and Lemay [7], and Jones, Nickels, and Marusic [8] have already developed flow governing approaches for the defect profile. The version below is different in that we use a stream function approach like that used in Section 2.1. To implement the defect velocity based *x*-momentum equation we will start out the same way as for the traditional approach used in Section 2.1 above. We use the same dimensionless independent variables from Eq. 2.2 as well as taking $\delta_s(x)$ as the yet unknown boundary layer thickness scaling parameter. Next we introduce a new stream function based on the defect velocity. Thus we assume that a stream function $\psi_d(x,y)$ exists such that

$$\psi_d(x,y) \;=\; \delta_s(x)u_e(x)\eta - \delta_s(x)u_s(x)g(\xi,\eta) \quad , \tag{2.17}$$

and that satisfies the conditions

$$u \;=\; \frac{\partial \psi_d(x,y)}{\partial y}, \qquad v \;=\; -\frac{\partial \psi_d(x,y)}{\partial x} \quad , \tag{2.18}$$

where $u_s(x)$ is the as yet unknown scaling velocity, $u_e(x)$ is the velocity at the boundary layer edge, and $g(\xi,\eta)$ is an unitless function.

Substituting Eq. 2.17 into Eq. 2.18, it is easily verified the defect profile is given by

$$u_e(x) - u(x,y) \;=\; u_s(x)g'(\xi,\eta) \quad . \tag{2.19}$$

Since the reduction of the x-momentum equation to a dimensionless equation is very tedious, it is moved to Appendix B. The result of transforming the *x*-component of the momentum balance is given as Eq. B.12 and repeated here as

$$\begin{aligned} &\cancel{u_e\frac{du_e}{dx}} - \left(u_e\frac{du_s}{dx} + u_s\frac{du_e}{dx}\right)g' + \left(\frac{u_e u_s}{\delta_s}\frac{d\delta_s}{dx} + u_s\frac{du_e}{dx}\right)\eta g'' + u_s\frac{du_s}{dx}g'^2 + \\ &-u_s\frac{du_s}{dx}gg'' + \frac{\partial}{\partial x}\left\{\overline{\tilde{u}^2} - \overline{\tilde{v}^2}\right\} + \frac{\partial}{\partial y}\left\{\overline{\tilde{u}\tilde{v}}\right\} \;=\; \cancel{u_e\frac{du_e}{dx}} - \nu\frac{u_s}{\delta_s^2}g''' \quad , \end{aligned} \tag{2.20}$$

where we have assumed $g(\xi,\eta)$ is now only a function of $\eta$.

In order to make the momentum equation dimensionless we will divide Eq. 2.20 through by $u_s \frac{du_s}{dx}$. Thus the first $x$-grouping is

$$\frac{u_e \frac{du_s}{dx} + u_s \frac{du_e}{dx}}{u_s \frac{du_s}{dx}} = \frac{u_e}{u_s} + \frac{\frac{du_e}{dx}}{\frac{du_s}{dx}} = \frac{2}{C_0} \quad , \tag{2.21}$$

where $C_0$ must be a constant for similarity. (The reason for using $C_0$, the same constant from Eq. 2.6, will be obvious shortly).

The next $x$-grouping term reduces to

$$\begin{aligned} \frac{\frac{u_e u_s}{\delta_s}\frac{d\delta_s}{dx} + u_s \frac{du_e}{dx}}{u_s \frac{du_s}{dx}} &= \frac{u_e \frac{d\delta_s}{dx}}{\delta_s \frac{du_s}{dx}} + \frac{2}{C_0} - \frac{u_e}{u_s} \\ &= -\frac{u_e}{u_s \Lambda_s} + \frac{2}{C_0} - \frac{u_e}{u_s} = C_1 \quad , \end{aligned} \tag{2.22}$$

where $\Lambda_s$ is from Eq. 2.7. $C_1$ must be constant for similarity.

For the last $x$-grouping term we will use the parameter $\beta$ from Eq. 2.9 so that

$$\frac{\nu \frac{u_s}{\delta_s^2}}{u_s \frac{du_s}{dx}} = \frac{\nu}{\delta_s^2 \frac{du_s}{dx}} = \frac{1}{\beta} \quad . \tag{2.23}$$

The transformed $x$-component of the momentum balance (Eq. 2.20) therefore reduces to

$$\begin{aligned} &\frac{1}{\beta} g''' - \frac{2}{C_0} g' + \left( -\frac{u_e}{u_s \Lambda_s} + \frac{2}{C_0} - \frac{u_e}{u_s} \right) \eta g'' - \left( 1 - \frac{1}{\Lambda_s} \right) g g'' + g'^2 + \\ &+ \{\text{Reynolds Stress Terms}\} = 0 \quad , \end{aligned} \tag{2.24}$$

where the Reynolds stress terms are left un-transformed since they can have no involvement with the behavior of $\delta_s$ or $u_s$ without invoking some closure assumption.

### 2.2.a Similarity of the Defect Profile

The $x$-grouping ratios given by Eqs. 2.21-2.23 constitute the similarity criteria. The three scaling constants $C_0$, $C_1$, and $\beta$ must be constant for similarity to be present. Of these three, only the $C_0$ derived from the pressure gradient $x$-grouping ratio involves both our unknown velocity scaling parameter $u_s(x)$ and the velocity at the boundary layer edge $u_e(x)$. For similar solutions it is a necessary condition that the $x$-groupings must be a constant. For $C_0$ equal to a constant, Eq. 2.21 has a solution given by

$$u_s(x) = \frac{C_0 u_e(x)}{2} \pm \sqrt{b + \frac{C_0^2 u_e^2(x)}{4}} \quad , \tag{2.25}$$

where *b* is a constant. There are no restrictions on the value for *b* so we take *b*=0 which then means that one of the nontrivial solutions is

$$u_s(x)/u_e(x) \ = \ C_0 \quad , \tag{2.26}$$

such that $C_0$ must be a constant for similarity to be present. This is the same similarity condition discovered using previously published flow governing equation approaches [2-3, 7-9]. It is of course the same equation as Eq. 2.15 above.

Using this equation, Eq. 2.22 reduces to

$$\begin{aligned} C_1 \ &= \ -\frac{u_e}{u_s\Lambda_s}+\frac{2}{C_0}-\frac{u_e}{u_s} \ = \ -\frac{1}{C_0\Lambda_s}+\frac{1}{C_0} \\ &= \ \frac{1}{C_0}\left(1-\frac{1}{\Lambda_s}\right) \ = \ \frac{\alpha}{C_0\beta} \quad , \end{aligned} \tag{2.27}$$

where $\alpha$ is from Eq. 2.11. Therefore the reduced flow governing equation for defect profile becomes

$$g'''+\alpha\left(\frac{\eta g''}{C_0}-gg''\right)-\beta\left(\frac{2}{C_0}g'-g'^2\right)+\{\text{Reynolds Stress Terms}\} \ = \ 0 \quad . \tag{2.28}$$

This equation is the defect profile equivalent of the velocity profile based Falkner-Skan Eq. 2.12. If the fluctuating components $\tilde{u}$ and $\tilde{v}$ are identically zero, then the Reynolds stress terms in our dimensionless momentum equation, Eq. 2.28, disappear. Couple that with taking $C_0 = 1$, then Eq. 2.28 reduces to

$$g'''+\alpha(\eta-g)g''-\beta\left(2g'-g'^2\right) \ = \ 0 \quad , \tag{2.29}$$

a Falkner-Skan [11] like similarity equation for laminar flow on a flat plate with a pressure gradient.

## 2.3 Defect Profile and Velocity Profile Comparison

It is important to note that we verified that the laminar flow velocities $u(x,y)$ and $v(x,y)$ calculated using Eq. 2.29 are identical to the traditional Falkner-Skan equation solution calculated with Eq. 2.16. The first verification was done numerically. A simple Fortran program to calculate the Falkner-Skan equation was modified to calculate solutions to Eq. 2.29. It was then a simple process to compare the two numerical solutions for a couple of different $\alpha$ and $\beta$ values. The results for $u(x,y)$ and $v(x,y)$ calculated with the two equations are numerically indistinguishable. As a further check between the traditional equation and the defect profile equation approaches, we can do the simple mathematical comparison starting by forcing the stream velocities to be the same. This is done in Appendix C below. Hence, it is confirmed both numerically and theoretically that the Falkner-Skan equation solutions using Eq. 2.29 are identical to the traditional Falkner-Skan equation solutions using Eq. 2.16. This equivalence gives us confidence that derivations given above were done correctly and that the resulting similarity constraints $\alpha$, $\beta$, and $C_0$ were also derived correctly for the two cases.

Now we are in the position of comparing the similarity criteria for the two cases. The three similarity parameters $\alpha$, $\beta$, and $C_0$ for the velocity profile dimensionless *x*-momentum equation given by Eq. 2.12 and the three similarity parameters for the defect profile dimensionless $x$-momentum equation given by Eq. 2.28 are IDENTICAL. Thus any similarity scaling parameter $u_s(x)$ that satisfies the defect profile similarity requirements must also satisfy the velocity profile similarity requirements.

## 3. Defect and Velocity Profile Similarity

In the last Section it was demonstrated that as far as similarity criteria is concerned, the defect profile and the velocity profile are identical. If this is true then how does one explain Figs. 2 and 3 from Clauser [1]? In that paper, Clauser took datasets from the literature and plotted them as velocity profiles (his Fig. 2) and defect profiles (his Fig. 3). The defect profiles showed similar behavior but the velocity profiles did not when both sets were scaled with the skin friction velocity and the boundary layer thickness. If the similarity criteria are identical then it would not seem possible to have similarity in one case but not the other. Recently Weyburne [5] observed this same behavior in other datasets taken from the literature that were used to claim similar behavior based on plots of the defect profile. After further investigation, Weyburne found that many of those datasets were cases of "false" similarity. To understand this false similarity scenario, we need to start with the traditional definition of similarity of the velocity profile given by Schlichting [9] together with a simple similarity equivalency derivation used by Weyburne [10]. Take the length scaling parameter as $\delta_s$ and velocity scaling parameter as $u_s$. These scaling parameters can vary with the flow direction ($x$-direction) but not in the direction perpendicular to the wall ($y$-direction). According to Schlichting, the scaled velocity profile at a station $x_1$ along the wall will be similar to the scaled profile at $x_2$ if

$$\frac{u(x_1, y/\delta_s)}{u_s(x_1)} = \frac{u(x_2, y/\delta_s)}{u_s(x_2)} \quad \text{for all y.} \tag{3.1}$$

Using the above notations, defect profile similarity would therefore be given by

$$\frac{u_e(x_1) - u(x_1, y/\delta_s)}{u_s(x_1)} = \frac{u_e(x_2) - u(x_2, y/\delta_s)}{u_s(x_2)} \quad \text{for all y.} \tag{3.2}$$

By inspection, one can see that the defect profile similarity will be equivalent to velocity profile similarity if

$$\frac{u_e(x_1)}{u_s(x_1)} = \frac{u_e(x_2)}{u_s(x_2)} \quad . \tag{3.3}$$

If the Eq. 3.3 criterion is satisfied, then the defect profiles and the velocity profile must show similarity simultaneously. However, note that Eq. 3.3 also is a similarity condition discovered for defect profile similarity in flow governing equation approaches developed by Rotta [2] (see his Eq. 14.3), Townsend [3] (see his Eq. 7.2.3), Castillo and George [6] (see their Eq. 9), Maciel, Rossignol, and Lemay [7] (see their Eq. 26), and Jones, Nickels, and Marusic [8] (see their Eq. 3.9 a2+a4). Not surprisingly, Eq. 3.3 is also one of the similarity criteria derived above for velocity profile similarity (Eq. 2.15) and for defect profile similarity (Eq. 2.26). Furthermore, in Appendix C we offer three additional theoretical derivations that indicate that $u_e / u_s$ = constant is required for similar behavior. This point is worth emphasizing; there are at least two different theoretical treatments that indicate that the constraint $u_e/u_s$ = constant must hold in order to observe similar behavior in a set of profiles. This requirement, in turn, means that defect profile similarity must be accompanied by velocity profile similarity. You cannot have one without the other.

The theory indicates that defect profile similarity must be accompanied by velocity profile similarity. However, Clauser, and more recently Weyburne [5], observed that turbulent boundary layer datasets could display similarity when plotted as scaled defect profiles but the same data and scaling parameters plotted as velocity profiles does not show similarity. This anomalous behavior triggered a further investigation which resulted in what Weyburne called "true" and "false" similarity. The false similarity case occurs when defect similarity is present but the similarity criteria $u_e/u_s$ = constant is not satisfied. To understand how this can happen we observe that although Eq. 3.2 is the formal definition of defect similarity, it is often written in the form

$$\frac{u_e(x)-u(x,y/\delta_s)}{u_s(x)} \;=\; f(y/\delta_s) \quad , \tag{3.4}$$

where $f$ is some universal profile of the dimensionless height. Now examination of the left-hand side reveals that there are two ways by which one can obtain similar behavior. If both ratios are constant on the left-hand side then the difference will be constant. This scenario is what Weyburne termed "true" similarity. A second way to make the left side invariant in $x$ is to have each ratio change with $x$ in such a way that the difference is constant. In this "false" similarity scenario, the set of defect profiles will appear to show similar behavior but the similarity criteria $u_e/u_s$ = constant is not satisfied. The easiest way to observe whether false similarity is present or not is to plot the data as both scaled defect profiles and scaled velocity profiles using the same scaling parameters. If the defect profile plot shows similarity but the velocity profile plot does not then this is a case of false similarity and overall similarity is not indicated. This behavior is in fact what was observed by Clauser and Weyburne. As noted by Weyburne, this false similarity scenario seems to be a widespread occurrence in the turbulent boundary layer literature.

## 4. Discussion

The search for similarity parameters for the turbulent boundary layer has a long history going back to the experimental and theoretical work of Clauser [1]. In fact Clauser [1] seems to be the one to start the use of the defect profile in the search for similarity in experimental datasets. His Figs. 2 and 3 compare a number of experimental datasets plotted as velocity profiles and defect profiles. In those plots the defect profiles from different groups collapsed to a single universal profile but the velocity profiles did not. Subsequent searches for similarity scaling parameters for the turbulent boundary layer have adapted the use of the defect profile as a means of "discovering" similar behavior. Following Clauser's work, Rotta [2] and Townsend [3] developed defect profile based theoretical treatments for the turbulent boundary layer similarity. Subsequently, other flow governing equation based theoretical treatments of the turbulent boundary layer have been developed by Castillo and George [6], Maciel, Rossignol, and Lemay [7], and Jones, Nickels, and Marusic [8]. This association of the defect profile with the turbulent boundary layer has been reinforced by the extensive work that has occurred on the turbulent boundary layers inner region. The foundation for the Logarithmic Law of the Wall that describes the velocity profile behavior in the near wall region is the von Kármán's [4] analytical expression of the defect profile.

The early work by Clauser, Rotta, and Townsend all took the stance that if one is to find similarity in the turbulent boundary layer then it must be the case that the velocity scaling parameter must be proportional to the inner regions velocity scale given by Prandtl's "Plus" friction velocity. At that time, they still were treating similarity as a property of the whole profile. As noted by Clauser [1], turbulent boundary layer similarity is difficult to achieve experimentally hence there were few experiments which showed true similarity. Subsequently, Castillo and George [6] took a different stance. They argued that the turbulent boundary could be considered to be made of separate inner and outer regions which are characterized by different scaling parameters. This division is highlighted by Castillo and George's flow governing equation theoretical treatment in which they specifically search for outer region similarity scaling parameters without reference to or connection with the inner region. Hence, the current paradigm is that the inner and outer regions not only have separate scaling parameters but whereas the inner region is characterized by similarity of the velocity profile (Log Law plots for example), the outer region is characterized by similarity of the defect profile.

Inherent in this separate defect profile similarity and velocity profile similarity idea is a subtle but profound implication as to the physics of this flow situation. The conventional thinking is that these different plots are "discovering" the true physics inherent in the flow. However, stepping back and looking at it from a different perspective; what this amounts to is that by a simple shift of the data (*i.e.*, plotting $u_e - u$ instead of $u$) that this somehow changes the physics of the flow. To be clear on this point; the defect profile at any point along the flow is not a measured experimental dataset, it is a simple calculated profile that is based solely on the experimentally measured velocity profile data taken at that measuring station. One can generate any number of calculated profiles from the measured experimental data. From this perspective, a simple mathematical rearrangement of the experimental data does not change the underlying physics. Hence the similarity scaling parameters should work equally well for the calculated profiles and the original experimental velocity profile. But experimentally, this is not what is observed by Clauser, for example. Given Clauser's early observation and the lack of any theoretical evidence to the contrary, the accepted paradigm has been that the defect profile is 'discovering' similarity behavior that the velocity profile did not find.

Although Clauser's experimental evidence indicates that that the defect profile is different and distinct from the velocity profile, there is no theoretical evidence/proof to this belief. In Section 2 above, we sought to uncover a theoretical basis for this difference/distinction. What we showed instead is that the similarity criteria are identical for the two profiles. Therefore, if a certain thickness scaling parameter $\delta_s(x)$ and a certain velocity scaling parameter $u_s(x)$ make the defect profile behave in a similar manner than those same scaling parameters must also work for the velocity profile. Furthermore, in Section 3 we showed that defect profile similarity must be accompanied by velocity profile similarity. The fact that this similarity equivalence has not been noticed before in the literature is astonishing. For more than sixty years beginning with Rotta [2] and Townsend [3], the similarity requirement that $u_e / u_s$ = constant has been known but the implications have not been appreciated. However, examination of Eqs. 3.1-3.3 makes it clear; if this constraint is satisfied then defect profile similarity must be accompanied by velocity profile similarity. You cannot have true similar behavior unless both profiles show similar behavior. The reason this equivalence was not noticed earlier seems to be the early observation by Clauser plus the large Log Law theory supporting use of the defect profile created an impression that the defect profile is a different animal than the velocity profile. Now we see from the work herein that this is not the case; the velocity profile similarity should work equally well for the turbulent boundary layer or the laminar boundary layer.

Some will argue that the similarity equivalence argument offered in Section 3 above is too restrictive because it is based on whole profile similarity which requires similarity in both the inner and outer regions. Castillo and George's [6] flow governing equation development was intentionally limited to the outer region of the boundary layer even though Castillo and George never defined exactly where the outer region boundary ends and the inner region begins. In the same spirit; examination of Eqs. 3.1-3.2 indicates that the only part of the equivalence argument that changes by restricting the argument to just the outer region is that instead of applying "for all y" in the general case it becomes "for all y in the outer region" for the case restricted to the outer boundary layer region. The requirement that $u_e/u_s$ = constant still applies based on Castillo and George's [6] own development for the outer region (see their Eq. 9) which means that defect profile similarity must be accompanied by velocity profile similarity even in the case where only the outer region is considered.

If there is no theoretical preference between the defect profile and the velocity profile then what about experimental preferences. The experimental based reason to use the scaled defect profile instead of the scaled velocity profile is also not well supported. What is true in experimental measurements of the velocity profile is that near-wall measurements are problematic. Hence most datasets available in the literature tend to consist mostly of outer region data. In the outer region, the defect profile goes to zero as the boundary layer edge is approached. Thus in the very outer region of the boundary layer the defect profile is not helping to see differences in the plotted profiles but is hiding differences instead. The scaled velocity profile is just the opposite. The largest differences for this case tend to occur in the outer region. Hence if one is truly interested in studying the outer region similarity, then one should be looking at the scaled velocity profiles instead of scaled defect profiles.

## 6. Conclusions

A flow governing equation approach to similarity was used to show that the derived similarity criteria for the defect profile are identical to the similarity criteria derived for the velocity profile. This means that similarity scaling parameters should work equally well for the defect profile as well as the velocity profile.

## Acknowledgement

The author would like to acknowledge the support of the Air Force Research Laboratory and Dr. Gernot Pomrenke at AFOSR.

## Appendix A: Velocity Profile Flow Governing Equation Approach

In this section the flow governing equation approach to developing similarity criterion utilizing the velocity profile is given in detail. The 2-D fluid flow past a plate can be modeled theoretically by a combination of the Reynolds-averaged Navier-Stokes equation and the continuity equation. Assume the *x*-axis is placed in the plane of the plate in the stream flow direction, the *y*-axis is at right angles to this plate, and the *z*-axis is along the apex of the plate. Furthermore, assume only steady state solutions are considered. We start with the Reynolds decomposition of the velocities into a mean (or average) component and a fluctuating component

$$\begin{aligned} \hat{u} &= u + \tilde{u} \\ \hat{v} &= v + \tilde{v} \end{aligned} \qquad \text{(A.1)}$$

where the average components are $u$, the *x*-velocity, and $v$, the *y*-velocity. The fluctuating components are $\tilde{u}$ and $\tilde{v}$.

For a two-dimensional, incompressible, turbulent boundary layer, the Prandtl boundary layer *x*-component of the momentum balance is given by

$$u\frac{\partial u}{\partial x} + v\frac{\partial u}{\partial y} + \frac{\partial}{\partial x}\left\{\overline{\tilde{u}^2} - \overline{\tilde{v}^2}\right\} + \frac{\partial}{\partial y}\left\{\overline{\tilde{u}\tilde{v}}\right\} \cong -\frac{1}{\rho}\frac{\partial p}{\partial x} + \nu\frac{\partial^2 u}{\partial y^2} \quad , \qquad \text{(A.2)}$$

where $\rho$ is the density, $p$ is the pressure, and $\nu$ is the kinematic viscosity.

### A.1 Traditional Stream Function and Variable Transform

In order to reduce the *x*-momentum equation to a dimensionless equation we introduce the dimensionless independent variables

$$\xi = C_\xi x \ , \quad \eta = \frac{y}{\delta_s(x)} \quad , \qquad \text{(A.3)}$$

where $C_\xi$ is a constant with units of 1/distance and the function $\delta_s(x)$ is an as yet unknown boundary layer thickness scaling parameter that is a function of *x*. Next we introduce a stream function so that the mass conservation equation will automatically be satisfied. Underlying this stream function approach is a critical assumption to the whole theoretical development and that is that the velocities $u(x,y)$ and $v(x,y)$ can be decomposed into a product of *x*-functionals and scaled *y*-functionals. Thus we assume that a stream function $\psi(x,y)$ exists such that

$$\psi(x,y) = \delta_s(x) u_s(x) f(\xi,\eta) \quad , \qquad \text{(A.4)}$$

where $u_s$(x) is the velocity scaling parameter, $f(\xi,\eta)$ is a dimensionless function, and the stream function satisfies the conditions

$$u(x,y) = \frac{\partial\psi(x,y)}{\partial y}, \qquad v(x,y) = -\frac{\partial\psi(x,y)}{\partial x} \quad . \qquad \text{(A.5)}$$

This means that

$$
\begin{aligned}
v &= -\frac{\partial \psi}{\partial x} \\
v &= -\frac{\partial}{\partial x}\left\{ \delta_s(x) u_s(x) f(\xi,\eta) \right\} \\
v &= -\frac{d\{\delta_s(x)u_s(x)\}}{dx} f - \delta_s(x)u_s(x)\frac{\partial f}{\partial x} \\
v &= -\frac{d\{\delta_s(x)u_s(x)\}}{dx} f - \delta_s(x)u_s(x)\left( \frac{\partial f}{\partial \eta}\frac{\partial \eta}{\partial x} + \frac{\partial f}{\partial \xi}\frac{\partial \xi}{\partial x} \right) \\
v &= -\frac{d\{\delta_s u_s\}}{dx} f - \delta_s u_s \left( -\frac{\eta}{\delta_s}\frac{d\delta_s}{dx} f' + C_\xi \frac{\partial f}{\partial \xi} \right) \\
v &= -\frac{d\{\delta_s u_s\}}{dx} f + u_s \frac{d\delta_s}{dx}\eta f' - \delta_s u_s C_\xi \frac{\partial f}{\partial \xi}
\end{aligned}
\tag{A.6}
$$

where the prime indicates differentiation with respect to $\eta$, and where we have used the fact that

$$
\frac{\partial \eta}{\partial x} = \frac{\partial}{\partial x}\left\{ \frac{y}{\delta_s(x)} \right\} = -\frac{\eta}{\delta_s}\frac{d\delta_s}{dx} \quad . \tag{A.7}
$$

We will also need

$$
\frac{\partial \eta}{\partial y} = \frac{\partial}{\partial y}\left\{ \frac{y}{\delta_s(x)} \right\} = \frac{1}{\delta_s} \quad , \tag{A.8}
$$

so that the scaled streamwise velocity becomes

$$
\begin{aligned}
u &= \frac{\partial \psi}{\partial y} = \frac{\partial\left\{ \delta_s(x) u_s(x) f(\xi,\eta) \right\}}{\partial y} \\
u &= \delta_s u_s \frac{\partial f}{\partial y} \\
u &= \delta_s u_s \frac{\partial f}{\partial \eta}\frac{\partial \eta}{\partial y} \\
u &= \delta_s u_s f' \frac{1}{\delta_s} \\
u &= u_s f' \quad ,
\end{aligned}
\tag{A.9}
$$

where we have used the fact that $\frac{\partial \xi}{\partial y} = 0$.

## A.2 Traditional Transformed Momentum Equation

Substituting the above similarity variables into the *x*-component of the momentum equation Eq. A.2; starting on the left-hand side, we have

$$
\begin{aligned}
u\frac{\partial u}{\partial x} &= u_s f' \frac{\partial \{u_s f'\}}{\partial x} \\
&= u_s f' \left\{ u_s \frac{\partial f'}{\partial x} + f' \frac{\partial u_s}{\partial x} \right\} \\
&= u_s f' \left\{ u_s \left( \frac{\partial f'}{\partial \xi}\frac{\partial \xi}{\partial x} + \frac{\partial f'}{\partial \eta}\frac{\partial \eta}{\partial x} \right) + f' \frac{du_s}{dx} \right\} \\
&= u_s f' \left\{ u_s \left( C_\xi \frac{\partial f'}{\partial \xi} - \frac{\eta}{\delta_s}\frac{d\delta_s}{dx} f'' \right) + f' \frac{du_s}{dx} \right\} \\
&= -\frac{u_s^2}{\delta_s}\frac{d\delta_s}{dx}\eta f' f'' + u_s \frac{du_s}{dx} f'^2 + u_s^2 f' C_\xi \frac{\partial f'}{\partial \xi} \quad .
\end{aligned}
\tag{A.10}
$$

The next term is

$$
\begin{aligned}
v\frac{\partial u}{\partial y} &= \left\{ -\frac{d\{\delta_s u_s\}}{dx} f + u_s \frac{d\delta_s}{dx}\eta f' - \delta_s u_s C_\xi \frac{\partial f}{\partial \xi} \right\} \frac{\partial \{u_s f'\}}{\partial \eta}\frac{\partial \eta}{\partial y} \\
&= \left\{ -\frac{d\{\delta_s u_s\}}{dx} f + u_s \frac{d\delta_s}{dx}\eta f' - \delta_s u_s C_\xi \frac{\partial f}{\partial \xi} \right\} u_s f'' \frac{1}{\delta_s} \\
&= \frac{u_s}{\delta_s}\left\{ -\frac{d\{\delta_s u_s\}}{dx} f f'' + u_s \frac{d\delta_s}{dx}\eta f' f'' - \delta_s u_s C_\xi \frac{\partial f}{\partial \xi} f'' \right\} \\
&= -\frac{u_s}{\delta_s}\frac{d\{\delta_s u_s\}}{dx} f f'' + \frac{u_s^2}{\delta_s}\frac{d\delta_s}{dx}\eta f' f'' - u_s^2 C_\xi f'' \frac{\partial f}{\partial \xi} \quad .
\end{aligned}
\tag{A.11}
$$

Combining these terms

$$
\begin{aligned}
\left\{ u\frac{\partial u}{\partial x} + v\frac{\partial u}{\partial y} \right\} &= \cancel{-\frac{u_s^2}{\delta_s}\frac{d\delta_s}{dx}\eta f' f''} + u_s \frac{du_s}{dx} f'^2 + u_s^2 f' C_\xi \frac{\partial f'}{\partial \xi} + \\
& -\frac{u_s}{\delta_s}\frac{d\{\delta_s u_s\}}{dx} f f'' + \cancel{\frac{u_s^2}{\delta_s}\frac{d\delta_s}{dx}\eta f' f''} - u_s^2 C_\xi f'' \frac{\partial f}{\partial \xi} \\
&= u_s \frac{du_s}{dx} f'^2 - \frac{u_s}{\delta_s}\frac{d\{\delta_s u_e\}}{dx} f f'' + u_s^2 f' C_\xi \frac{\partial f'}{\partial \xi} - u_s^2 C_\xi f'' \frac{\partial f}{\partial \xi} \quad .
\end{aligned}
\tag{A.12}
$$

The next step is to transform the viscous component in Eq. A.2 given by

$$\nu\frac{\partial^2 u}{\partial y^2} = \nu\frac{\partial\left\{\frac{\partial\{u_s f'\}}{\partial\eta}\frac{\partial\eta}{\partial y}\right\}}{\partial\eta}\frac{\partial\eta}{\partial y}$$

$$= \nu\frac{\partial\left\{\frac{\partial\{u_s f'\}}{\partial\eta}\frac{1}{\delta_s}\right\}}{\partial\eta}\frac{1}{\delta_s} \tag{A.13}$$

$$= \nu\frac{u_s}{\delta_s^2}f''' \quad .$$

The Euler (Bernoulli) equation is used for the pressure term $dp/dx$ so that

$$-\frac{1}{\rho}\frac{dp}{dx} = u_e\frac{du_e}{dx} \quad , \tag{A.14}$$

where $u_e(x)$ is the velocity at the boundary layer edge.

Combining the transformed terms, we get the transformed *x*-component of the momentum balance as

$$u_s\frac{du_s}{dx}f'^2 - \frac{u_s}{\delta_s}\frac{d\{\delta_s u_s\}}{dx}ff'' + u_s^2 C_\xi f'\frac{\partial f'}{\partial\xi} - \delta_s u_s C_\xi f''\frac{\partial f}{\partial\xi} + \tag{A.15}$$

$$+\{\text{Reynolds Stress Terms}\} = u_e\frac{du_e}{dx} + \nu\frac{u_s}{\delta_s^2}f''' \quad .$$

Recall that similarity of the velocity profile for 2-D wall-bounded flows is defined as the case where two velocity profiles taken from different stations along the flow differ only by simple scaling parameters in $y$ and $u(x,y)$. In order to see this behavior in Eq. A.15, we start by assuming that the functional $f(\xi,\eta)$ can only depend on the scaled *y*-variable $\eta$. This means that the terms involving differentiation with respect to $\xi$ must be zero. Furthermore, similarity requires that each *x*-variable grouping must change proportionately as we move from station to station. Alternatively, we can divide through by one of the *x*-variable groupings which changes the similarity condition to be that each x-variable grouping ratio must be constant. Using the latter approach, we divide Eq. A.15 by $u_s\,du_s/dx$. Thus the pressure gradient *x*-grouping ratio becomes

$$\frac{u_e\frac{du_e}{dx}}{u_s\frac{du_s}{dx}} = \frac{1}{C_0^2} \quad , \tag{A.16}$$

where $C_0$ must be a constant for similarity. Let

$$\Lambda_s = -\frac{\delta_s}{u_s}\frac{du_s/dx}{d\delta_s/dx} \quad , \tag{A.17}$$

and use this in the next *x*-grouping term given by

$$\frac{\dfrac{u_s}{\delta_s}\dfrac{d\{\delta_s u_s\}}{dx}}{u_s\dfrac{du_s}{dx}} = \frac{u_s\dfrac{du_s}{dx}+\dfrac{u_s^2}{\delta_s}\dfrac{d\delta_s}{dx}}{u_s\dfrac{du_s}{dx}} = 1+\frac{\dfrac{u_s^2}{\delta_s}\dfrac{d\delta_s}{dx}}{u_s\dfrac{du_s}{dx}} = 1+\frac{u_s\dfrac{d\delta_s}{dx}}{\delta_s\dfrac{du_s}{dx}} = 1-\frac{1}{\Lambda_s} \quad . \tag{A.18}$$

Hence $\Lambda_s$ must be a constant for similarity. The viscous $x$-grouping is

$$\frac{\nu\dfrac{u_s}{\delta_s^2}}{u_s\dfrac{du_s}{dx}} = \frac{\nu}{\delta_s^2\dfrac{du_s}{dx}} = \frac{1}{\beta} \quad , \tag{A.19}$$

where $\beta$ must be a constant for similarity. Substituting into Eq. A.15 we get the transformed $x$-component of the momentum balance as

$$\frac{f'''}{\beta}+\left(1-\frac{1}{\Lambda_s}\right)ff''+\frac{1}{C_0^2}-f'^2+\{\text{Reynolds Stress Terms}\} = 0 \quad , \tag{A.20}$$

$$f'''+\beta\left(1-\frac{1}{\Lambda_s}\right)ff''+\beta\left(\frac{1}{C_0^2}-f'^2\right)+\{\text{Reynolds Stress Terms}\} = 0 \quad ,$$

where we have neglected terms involving $\xi$. We are leaving the Reynolds stress terms as unevaluated due to the fact that at the present time there is no implemented closure for turbulent boundary layer flows. Hence these Reynolds stress terms cannot provide guidance for the velocity profile behavior.

The equation begins to take on the appearance of the familiar Falkner-Skan equation, as it should. To emphasize this we introduce $\alpha$ so that

$$\alpha = \beta\left(1-\frac{1}{\Lambda_s}\right) = \frac{\delta_s^2\dfrac{du_s}{dx}}{\nu}\left(1+\frac{u_s\dfrac{d\delta_s}{dx}}{\delta_s\dfrac{du_s}{dx}}\right) \tag{A.21}$$

$$\alpha = \frac{\delta_s^2\dfrac{du_s}{dx}}{\nu}+\frac{\delta_s u_s\dfrac{d\delta_s}{dx}}{\nu} \quad .$$

Substituting this parameter into Eq. A.20, the dimensionless momentum equation becomes

$$f'''+\alpha ff''+\beta\left(\frac{1}{C_0^2}-f'^2\right)+\{\text{Reynolds Stress Terms}\} = 0 \quad . \tag{A.22}$$

## A.3 Velocity Profile Similarity Criteria

The $x$-grouping ratios given by Eqs. A.16, A.18, and A.19 constitute the similarity criteria. Similarity requires that the three parameters $C_0$, $\Lambda_s$, and $\beta$ must be constant otherwise the solutions obtained at each station would be different. Of these three, only the $C_0$ derived from the pressure gradient $x$-grouping ratio involves both our unknown velocity scaling parameter $u_s(x)$ and the velocity at the boundary layer edge $u_e(x)$. We can write Eq. A.16 as

$$u_s \frac{du_s}{dx} = C_0^2 u_e \frac{du_e}{dx} \quad , \tag{A.23}$$

where $C_0^2$ is a constant. This differential equation has a solution given by

$$u_s(x) = \sqrt{a + u_e^2(x) C_0^2} \quad , \tag{A.24}$$

where *a* is a constant. There is no restrictions on the value of *a* so we take *a*=0 which then means that one of the non-trivial solutions is given by

$$u_s(x)/u_e(x) = C_0 \quad . \tag{A.25}$$

This is one of the similarity conditions discovered by Rotta [2], Townsend [3], Castillo and George [6], Maciel, Rossignol, and Lemay [7], and Jones, Nickels, and Marusic [8] using a defect profile approach.

It is important to note that Eq. A.25 criteria is only obtained for the flow governing equation approach case in which the velocity profile boundary layer has a pressure gradient in the flow direction. For the zero pressure gradient (ZPG) case, Eq. A.16 goes away which means Eq. A.25 also goes away. The ZPG case should reduce to the Blasius [12] solution. The derivation requires that $u_s(x) = u_e(x)$ (see Appendix D) and that $du_e/dx = 0$. These requirements become the similarity criteria for the ZPG velocity profile case.

If the fluctuating components $\tilde{u}$ and $\tilde{v}$ are identically zero, then the Reynolds stress terms in our dimensionless momentum equation, Eq. A.22, disappear. Couple that with taking *a*=0 and $C_0 = 1$ in Eq. A.24, then Eq. A.22 reduces to

$$f''' + \alpha f f'' + \beta\left(1 - f'^2\right) = 0 \quad , \tag{A.26}$$

which is the Falkner-Skan similarity equation for laminar flow on a flat plate with a pressure gradient. In this case Eq. A.26, and $\alpha$ and $\beta$ terms (Eqs. A.21 and A.19), are identical to Schlichting's [9] version of the Falkner-Skan equation. Taking *a*=0 and $C_0 = 1$ in Eq. A.24 is consistent with the original laminar flow Falkner-Skan equation which assumes that $u_s(x) = u_e(x)$ right from the start.

## Appendix B: Defect Profile Flow Governing Equation Approach

In this section the flow governing equation approach to developing similarity criterion utilizing the defect profile is given in detail. Rotta [2], Townsend [3], Castillo and George [6], Maciel, Rossignol, and Lemay[7], and Jones, Nickels, and Marusic [8] have already developed flow governing approaches for the defect profile. The version below is different in that we use the stream function approach used in Section 3.1. To implement the defect velocity based *x*-momentum equation we will start out the same way as for the traditional approach used in Appendix B above. We use the same dimensionless independent variables from Eq. 3.2 (Eq. A.2) as well as taking $\delta_s(x)$ as the yet unknown boundary layer thickness scaling parameter. Next we introduce a new stream function based on the defect velocity. Thus we assume that a stream function $\psi_d(x,y)$ exists such that

$$\psi_d(x,y) = \delta_s(x) u_e(x) \eta - \delta_s(x) u_s(x) g(\xi,\eta) \quad , \tag{B.1}$$

and that satisfies the conditions

$$u(x,y) \;=\; \frac{\partial \psi_d(x,y)}{\partial y}, \qquad v(x,y) \;=\; -\frac{\partial \psi_d(x,y)}{\partial x} \quad . \tag{B.2}$$

where $u_s(x)$ is the as yet unknown scaling velocity, $u_e(x)$ is the velocity at the boundary layer edge, and $g(\xi,\eta)$ is an unitless function. This means that

$$\begin{aligned}
v &= -\frac{\partial \psi_d}{\partial x} = -\frac{\partial}{\partial x}\left\{ \delta_s(x)u_e(x)\eta - \delta_s(x)u_s(x)g(\xi,\eta)\right\} \qquad (B.3)\\
v &= -\frac{d\{\delta_s u_e\}}{dx}\eta - \delta_s u_e \frac{\partial \eta}{\partial x} + \frac{d\{\delta_s u_s\}}{dx} g + \delta_s u_s \frac{\partial g}{\partial x}\\
v &= -\frac{d\{\delta_s u_e\}}{dx}\eta - \delta_s u_e \left(-\frac{\eta}{\delta_s}\frac{d\delta_s}{dx}\right) + \frac{d\{\delta_s u_s\}}{dx} g + \delta_s u_s \left(\frac{\partial g}{\partial \eta}\frac{\partial \eta}{\partial x} + \frac{\partial g}{\partial \xi}\frac{\partial \xi}{\partial x}\right)\\
v &= \cancel{-u_e \frac{d\delta_s}{dx}}\eta - \delta_s \frac{du_e}{dx}\eta + \cancel{u_e \frac{d\delta_s}{dx}}\eta + \frac{d\{\delta_s u_s\}}{dx} g + \delta_s u_s \left(-\frac{\eta}{\delta_s}\frac{d\delta_s}{dx} g' + C_\xi \frac{\partial g}{\partial \xi}\right)\\
v &= -\delta_s \frac{du_e}{dx}\eta + \frac{d\{\delta_s u_s\}}{dx} g - u_s \frac{d\delta_s}{dx}\eta g' + \delta_s u_s C_\xi \frac{\partial g}{\partial \xi} \quad ,
\end{aligned}$$

where the prime indicates differentiation with respect to $\eta$. Using Eqs. A.7 and A.8, it is easily verified that the streamwise velocity is given by

$$\begin{aligned}
u &= \frac{\partial \psi_d}{\partial y} = \frac{\partial\left\{ \delta_s(x)u_e(x)\eta - \delta_s(x)u_s(x)g(\xi,\eta)\right\}}{\partial y} \qquad (B.4)\\
u &= \delta_s u_e \frac{\partial \eta}{\partial y} - \delta_s u_s \frac{\partial g}{\partial y}\\
u &= \delta_s u_e \frac{1}{\delta_s} - \delta_s u_s \frac{\partial g}{\partial \eta}\frac{\partial \eta}{\partial y}\\
u &= u_e - \delta_s u_s g' \frac{1}{\delta_s}\\
u &= u_e - u_s g' \quad ,
\end{aligned}$$

This is off course the velocity deficit profile which is more formally written as

$$u_e(x) - u(x,y) \;=\; u_s(x)\, g'(\xi,\eta) \quad . \tag{B.5}$$

### B.1 Transformed Defect Momentum Equation

Substituting the above reduced velocities into the *x*-component of the momentum equation, layer *x*-component of the momentum balance given by

$$u\frac{\partial u}{\partial x} + v\frac{\partial u}{\partial y} + \frac{\partial}{\partial x}\left\{\overline{\tilde{u}^2} - \overline{\tilde{v}^2}\right\} + \frac{\partial}{\partial y}\left\{\overline{\tilde{u}\tilde{v}}\right\} \;\cong\; -\frac{1}{\rho}\frac{\partial p}{\partial x} + \nu\frac{\partial^2 u}{\partial y^2} \quad , \tag{B.6}$$

we have, starting on the left-hand side,

$$u\frac{\partial u}{\partial x} \;=\; \{u_e - u_s g'\}\frac{\partial\{u_e - u_s g'\}}{\partial x} \tag{B.7}$$

$$=\{u_e - u_s g'\}\left\{\frac{du_e}{dx} - \frac{du_s}{dx}g' - u_s\frac{dg'}{dx}\right\}$$

$$=\{u_e - u_s g'\}\left\{\frac{du_e}{dx} - \frac{du_s}{dx}g' - u_s\frac{dg'}{d\eta}\frac{d\eta}{dx} - u_s\frac{dg'}{d\xi}\frac{d\xi}{dx}\right\}$$

$$=\{u_e - u_s g'\}\left\{\frac{du_e}{dx} - \frac{du_s}{dx}g' + \frac{u_s}{\delta_s}\frac{d\delta_s}{dx}\eta g'' - u_s C_\xi\frac{\partial g'}{\partial \xi}\right\}$$

$$=u_e\frac{du_e}{dx} - u_e\frac{du_s}{dx}g' + \frac{u_e u_s}{\delta_s}\frac{d\delta_s}{dx}\eta g'' - u_e u_s C_\xi\frac{\partial g'}{\partial \xi} +$$

$$-u_s\frac{du_e}{dx}g' + u_s\frac{du_s}{dx}g'^2 - \frac{u_s^2}{\delta_s}\frac{d\delta_s}{dx}\eta g' g'' + u_s^2 C_\xi g'\frac{\partial g'}{\partial \xi} \quad .$$

The next inertial term is

$$v\frac{\partial u}{\partial y} = \left\{-\delta_s\frac{du_e}{dx}\eta + \frac{d\{\delta_s u_s\}}{dx}g - u_s\frac{d\delta_s}{dx}\eta g' + \delta_s u_s C_\xi\frac{\partial g}{\partial \xi}\right\}\frac{\partial\{u_e - u_s g'\}}{\partial \eta}\frac{\partial \eta}{\partial y} \tag{B.8}$$

$$=-\left\{-\delta_s\frac{du_e}{dx}\eta + \frac{d\{\delta_s u_s\}}{dx}g - u_s\frac{d\delta_s}{dx}\eta g' + \delta_s u_s C_\xi\frac{\partial g}{\partial \xi}\right\}u_s g''\frac{1}{\delta_s}$$

$$=-\frac{u_s}{\delta_s}\left\{-\delta_s\frac{du_e}{dx}\eta g'' + \frac{d\{\delta_s u_s\}}{dx}g g'' - u_s\frac{d\delta_s}{dx}\eta g' g'' + \delta_s u_s C_\xi g''\frac{\partial g}{\partial \xi}\right\}$$

$$=u_s\frac{du_e}{dx}\eta g'' - \frac{u_s}{\delta_s}\frac{d\{\delta_s u_s\}}{dx}g g'' + \frac{u_s^2}{\delta_s}\frac{d\delta_s}{dx}\eta g' g'' - u_s^2 C_\xi g''\frac{\partial g}{\partial \xi} \quad .$$

Combining these two terms we have

$$v\frac{\partial u}{\partial y} + u\frac{\partial u}{\partial x} \;=\; u_e\frac{du_e}{dx} - u_e\frac{du_s}{dx}g' + \frac{u_e u_s}{\delta_s}\frac{d\delta_s}{dx}\eta g'' - u_e u_s C_\xi\frac{\partial g'}{\partial \xi} + \tag{B.9}$$

$$-u_s\frac{du_e}{dx}g' + u_s\frac{du_s}{dx}g'^2 \cancel{-\frac{u_s^2}{\delta_s}\frac{d\delta_s}{dx}\eta g' g''} + u_s^2 C_\xi g'\frac{\partial g'}{\partial \xi} +$$

$$u_s\frac{du_e}{dx}\eta g'' - \frac{u_s}{\delta_s}\frac{d\{\delta_s u_s\}}{dx}g g'' + \cancel{\frac{u_s^2}{\delta_s}\frac{d\delta_s}{dx}\eta g' g''} - u_s^2 C_\xi g''\frac{\partial g}{\partial \xi}$$

$$=u_e\frac{du_e}{dx} - \left(u_e\frac{du_s}{dx} + u_s\frac{du_e}{dx}\right)g' + \left(\frac{u_e u_s}{\delta_s}\frac{d\delta_s}{dx} + u_s\frac{du_e}{dx}\right)\eta g'' + u_s\frac{du_s}{dx}g'^2 +$$

$$-\left(\frac{u_s^2}{\delta_s}\frac{d\delta_s}{dx} + u_s\frac{du_s}{dx}\right)g g'' + u_s^2 C_\xi g'\frac{\partial g'}{\partial \xi} - u_s^2 C_\xi g''\frac{\partial g}{\partial \xi} \quad .$$

The next step is to transform the viscous component in Eq. B.6 so that

$$\nu\frac{\partial^2 u}{\partial y^2} = \nu\frac{\partial\left\{\frac{\partial\{u_e - u_s g'\}}{\partial\eta}\frac{\partial\eta}{\partial y}\right\}}{\partial\eta}\frac{\partial\eta}{\partial y} \tag{B.10}$$

$$= -\nu\frac{\partial\left\{\frac{\partial\{u_s g'\}}{\partial\eta}\frac{1}{\delta_s}\right\}}{\partial\eta}\frac{1}{\delta_s}$$

$$= -\nu\frac{u_s}{\delta_s^2}g''' \quad .$$

The Euler (Bernoulli) equation is used for the pressure term $dp/dx$ which means

$$-\frac{1}{\rho}\frac{dp}{dx} = u_e\frac{du_e}{dx} \quad . \tag{B.11}$$

Combining the transformed terms, we get the $x$-component of the momentum balance as

$$\cancel{u_e\frac{du_e}{dx}} - \left(u_e\frac{du_s}{dx} + u_s\frac{du_e}{dx}\right)g' + \left(\frac{u_e u_s}{\delta_s}\frac{d\delta_s}{dx} + u_s\frac{du_e}{dx}\right)\eta g'' + u_s\frac{du_s}{dx}g'^2 + \tag{B.12}$$

$$-\left(\frac{u_s^2}{\delta_s}\frac{d\delta_s}{dx} + u_s\frac{du_s}{dx}\right)gg'' + \frac{\partial}{\partial x}\left\{\overline{\tilde{u}^2} - \overline{\tilde{v}^2}\right\} + \frac{\partial}{\partial y}\left\{\overline{\tilde{u}\tilde{v}}\right\} = \cancel{u_e\frac{du_e}{dx}} - \nu\frac{u_s}{\delta_s^2}g''' \quad ,$$

where we have taken the step of neglecting terms involving $\xi$ as explained in Appendix A.

In order to make the momentum equation dimensionless we will divide Eq. B.12 through by $u_s\frac{du_s}{dx}$. Thus the first $x$-grouping is

$$\frac{u_e\frac{du_s}{dx} + u_s\frac{du_e}{dx}}{u_s\frac{du_s}{dx}} = \frac{u_e}{u_s} + \frac{\frac{du_e}{dx}}{\frac{du_s}{dx}} = \frac{2}{C_0} \quad , \tag{B.13}$$

where $C_0$ must be a constant for similarity. (The reason for using $C_0$, the same constant from Eq. A.16, will be obvious shortly).

The next $x$-grouping term reduces to

$$\frac{\frac{u_e u_s}{\delta_s}\frac{d\delta_s}{dx} + u_s\frac{du_e}{dx}}{u_s\frac{du_s}{dx}} = \frac{u_e\frac{d\delta_s}{dx}}{\delta_s\frac{du_s}{dx}} + \frac{2}{C_0} - \frac{u_e}{u_s} \tag{B.14}$$

$$= -\frac{u_e}{u_s\Lambda_s} + \frac{2}{C_0} - \frac{u_e}{u_s} = C_1 \quad ,$$

where $\Lambda_s$ is from Eq. A.17. $C_1$ must be a constant for similarity.

The next $x$-grouping term reduces to

$$\frac{\dfrac{u_s^2}{\delta_s}\dfrac{d\delta_s}{dx}+u_s\dfrac{du_s}{dx}}{u_s\dfrac{du_s}{dx}} = \frac{u_s\dfrac{d\delta_s}{dx}}{\delta_s\dfrac{du_s}{dx}}+1 = 1-\frac{1}{\Lambda_s} \quad , \tag{B.15}$$

For the last $x$-grouping term we will use the parameter $\beta$ from Eq. A.19 so that

$$\frac{\nu\dfrac{u_s}{\delta_s^2}}{u_s\dfrac{du_s}{dx}} = \frac{\nu}{\delta_s^2\dfrac{du_s}{dx}} = \frac{1}{\beta} \quad . \tag{B.16}$$

The transformed $x$-component of the momentum balance (Eq. B.12) therefore reduces to

$$\frac{1}{\beta}g''' - \frac{2}{C_0}g' + \left(-\frac{u_e}{u_s\Lambda_s}+\frac{2}{C_0}-\frac{u_e}{u_s}\right)\eta g'' - \left(1-\frac{1}{\Lambda_s}\right)gg'' + g'^2 + \tag{B.17}$$
$$+\{\text{Reynolds Stress Terms}\} = 0 \quad .$$

The Reynolds stress terms are left un-transformed since they can have no involvement with the behavior of $\delta_s$ or $u_s$ without invoking some closure assumption.

### B.2 Similarity of the Defect Profile

Similarity criteria can be derived from the x-grouping ratios given by Eqs. B.13-B.16. The similarity constants are $C_0$, $C_1$, $\Lambda_s$, and $\beta$. Of these, only the $C_0$ derived from the pressure gradient $x$-grouping ratio involves both our unknown velocity scaling parameter $u_s(x)$ and the velocity at the boundary layer edge $u_e(x)$. For similar solutions it is a necessary condition that the $x$-groupings must be a constant. For $C_0$=constant, Eq. B.13 has a solution given by

$$u_s(x) = \frac{C_0 u_e(x)}{2} \pm \sqrt{b+\frac{C_0^2 u_e^2(x)}{4}} \quad , \tag{B.18}$$

where $b$ is a constant. There are no restrictions on the value for $b$ so we take $b$=0 which then means that one of the non-trivial solutions is given by

$$u_s(x)/u_e(x) = C_0 \quad , \tag{B.19}$$

such that $C_0$ must be a constant for similarity to be present. This is the first similarity condition discovered by Rotta [2], Townsend [3], Castillo and George [6], Maciel, Rossignol, and Lemay [7], and Jones, Nickels, and Marusic [8]. It is of course the same equation as Eq. A.25 above.

Using this equation, Eq. B.14 reduces to

$$C_1 = -\frac{u_e}{u_s\Lambda_s}+\frac{2}{C_0}-\frac{u_e}{u_s} = -\frac{1}{C_0\Lambda_s}+\frac{1}{C_0} \tag{B.20}$$
$$= \frac{1}{C_0}\left(1-\frac{1}{\Lambda_s}\right) = \frac{\alpha}{C_0\beta} \quad ,$$

where $\alpha$ is from Eq. A.21. Therefore the reduced flow governing equation, Eq. B.17, for the defect profile becomes

$$\frac{1}{\beta}g''' - \frac{2}{C_0}g' + \frac{\alpha}{C_0\beta}\eta g'' - \left(1 - \frac{1}{\Lambda_s}\right)gg'' + g'^2 + \{\text{Reynolds Stress Terms}\} = 0, \tag{B.21}$$

$$g''' - \frac{2\beta}{C_0}g' + \frac{\alpha}{C_0}\eta g'' - \beta\left(1 - \frac{1}{\Lambda_s}\right)gg'' + \beta g'^2 + \{\text{Reynolds Stress Terms}\} = 0,$$

$$g''' - \frac{2\beta}{C_0}g' + \frac{\alpha}{C_0}\eta g'' - \alpha gg'' + \beta g'^2 + \{\text{Reynolds Stress Terms}\} = 0,$$

$$g''' + \alpha\left(\frac{\eta g''}{C_0} - gg''\right) - \beta\left(\frac{2}{C_0}g' - g'^2\right) + \{\text{Reynolds Stress Terms}\} = 0 \quad ,$$

The three constant parameters $\alpha$, $\beta$, and $C_0$ for the velocity profile dimensionless *x*-momentum equation given by Eq. A.22 and the three constant parameters for the defect profile dimensionless *x*-momentum equation given in Eq. B.21 are IDENTICAL. Hence the similarity criteria are identical. Thus any velocity scaling parameter that satisfies the defect profile must also satisfy the velocity profile.

If the fluctuating components $\tilde{u}$ and $\tilde{v}$ are identically zero, then the Reynolds stress terms in our dimensionless momentum equation, Eq. B.21, disappear. Couple that with taking $C_0 = 1$, then Eq. B.21 reduces to

$$g''' + \alpha(\eta - g)g'' - \beta\left(2g' - g'^2\right) = 0 \quad , \tag{B.22}$$

a Falkner-Skan like similarity equation for laminar flow on a flat plate with a pressure gradient. It is important to note that we verified that the calculated velocities $u(x,y)$ and $v(x,y)$ for Eq. B.22 are identical to the traditional Falkner-Skan equation solution Eq. A.26. The first verification was done numerically. A simple Fortran program to calculate the Falkner-Skan equation was modified to calculate solutions to Eq. B.22. It was then a simple process to compare the two numerical solutions for a couple of different $\alpha$ and $\beta$ values. The results were numerically indistinguishable. The second method was a simple substitution approach which is detailed in Appendix D.

**Appendix C: Alternative Derivation Check**

In order to show the equivalence of the traditional Falkner-Skan equation (Eq. A.26) and the defect profile based Falkner-Skan equation (Eq. B.22), we can do the simple mathematical comparison starting by forcing the stream velocities to be the same so that

$$
\begin{aligned}
u_e(x) - u_s(x) f'(\eta) &= u_s(x) g'(\eta) \\
g'(\eta) &= \frac{u_e(x) - u_s(x) f'(\eta)}{u_s(x)} \\
g'(\eta) &= 1 - f'(\eta),
\end{aligned} \tag{C.1}
$$

where we have assumed $u_s(x)$=$u_e(x)$. Integrating this equation we have

$$
\begin{aligned}
\int_0^\eta d\eta\, g'(\eta) &= \int_0^\eta d\eta \left(1 - f'(\eta)\right) \\
g(\eta) - g(0) &= \eta - f(\eta) - f(0) \\
g(\eta) &= \eta - f(\eta) \\
f(\eta) &= \eta - g(\eta) \quad .
\end{aligned} \tag{C.2}
$$

Differentiating Eq. C.1 by $\eta$ gives

$$
g''(\eta) = -f''(\eta) \quad . \tag{C.3}
$$

And differentiating a second time

$$
g'''(\eta) = -f'''(\eta) \quad . \tag{C.4}
$$

Combining into the Falkner-Skan equation, Eq. A.26, we have

$$
\begin{aligned}
f''' + \alpha f f'' + \beta\left(1 - f'^2\right) &= 0 \\
-g''' - \alpha(\eta - g) g'' + \beta\left(1 - (1 - g')^2\right) &= 0 \\
g''' + \alpha(\eta - g) g'' - \beta\left(1 - 1 + 2g' - g'^2\right) &= 0 \\
g''' + \alpha(\eta - g) g'' - \beta\left(2g' - g'^2\right) &= 0 \quad ,
\end{aligned} \tag{C.5}
$$

which is identical to Eq. B.22. Therefore it is confirmed that the two Falkner-Skan equations are equivalent.

## Appendix D: Alternative Derivations of the Velocity Ratio Scaling Criterion

The flow governing approaches to similarity by Rotta [2] (see his Eq. 14.3), Townsend [3] (see his Eq. 7.2.3), and Castillo and George [6] (see their Eq. 9), Maciel, Rossignol, and Lemay [7] (see their Eq. 26), and Jones, Nickels, and Marusic [8] (see their Eq. 3.9, a2+a4) all have as a similarity requirement that $u_e / u_s$=constant (or equivalent). The flow governing approach is not the only theoretical route to this criterion. We can offer at least three additional simple theoretical routes to substantiate $u_e / u_s$=constant as a valid similarity constraint.

First, start with the definition of defect similarity given by

$$
\frac{u_e(x) - u(x, y/\delta_s)}{u_s(x)} = f(y/\delta_s) \quad \text{for all y,} \tag{D.1}
$$

where $f$ is the universal profile function of the dimensionless height. Note that the "for all $y$" part is usually assumed. Since this equation must hold for all $y$; take the case where $y \to 0$ so that Eq. D.1 becomes

$$\frac{u_e(x)}{u_s(x)} = f(0) \quad , \tag{D.2}$$

where $f(0)$ is a constant. Therefore, $u_e/u_s$=constant is a valid similarity constraint.

Now let us consider a second theoretical route. Weyburne [14] recently used rigorous mathematics to prove that if similarity exists in a set of boundary layer profiles, then the displacement thickness $\delta_1$ must be a similar length scale and $u_e$ must be a similar velocity scale. This derivation starts with the traditional definition of similarity given according to Schlichting [9] as

$$\frac{u(x_1, y/\delta_s(x_1))}{u_s(x_1)} = \frac{u(x_2, y/\delta_s(x_2))}{u_s(x_2)} \quad \text{for all y.} \tag{D.3}$$

If similarity is present in a set of velocity profiles then it is self-evident that the properly scaled first derivative profile curves (derivative with respect to the scaled $y$-coordinate) must also be similar. It is also self-evident that the areas under the scaled first derivative profiles plotted against the scaled $y$-coordinate must be equal for similarity. In mathematical terms, the area under the scaled first derivative profile curve is expressed by

$$a(x) = \int_0^{h/\delta_s} d\left\{\frac{y}{\delta_s}\right\} \frac{d\left\{u(x, y/\delta_s)/u_s\right\}}{d\left\{\dfrac{y}{\delta_s}\right\}} \quad , \tag{D.4}$$

where $a(x)$ is in general a non-zero numerical value and $y = h$ is located deep in the free stream. With a variable switch $\left(d\{y/\delta_s\} \Rightarrow (1/\delta_s)dy\right)$, Eq. D.4 can be shown to reduce to

$$a(x) = \frac{u_e(x)}{u_s(x)} \quad . \tag{D.5}$$

Similarity requires that $a(x_1) = a(x_2) = \text{constant}$. Once again, $u_e/u_s$=constant must be a valid similarity constraint.

A third theoretical route also starts with Eq. D.3. Since this equation must hold for all $y$; take the case where $y \to h(x)$ such that $u(x, h(x)) = u_e(x)$. Similarity assumes that we end at the same scaled $y$-value given by

$$\frac{h(x_1)}{\delta_s(x_1)} = \frac{h(x_2)}{\delta_s(x_2)} \quad , \tag{D.6}$$

but this is easily satisfied since we can freely choose $h(x_1)$ and $h(x_2)$ to be located anywhere in the boundary layer edge region. If we choose the $y$-values $h(x_1)$ and $h(x_2)$ to satisfy Eq. D.6 at the boundary layer edge, then Eq. D.3 reduces to

$$\frac{u_e(x_1)}{u_s(x_1)} = \frac{u_e(x_2)}{u_s(x_2)} \quad . \tag{D.7}$$

Therefore we have three related theoretical formulations, all of which indicate that while other velocity scaling parameters are not excluded, it must be the case that if similarity exists in a set of velocity profiles then the velocity at the boundary layer edge $u_e$ must be a similarity scaling parameter for the 2-D boundary layer. Notice that none of these derivations are based on the flow governing equations but rather on the definition of similarity itself. Hence these arguments apply to any 2-D boundary layer flow that shows similarity from station to station.